\documentclass[twocolumn,final,twoside,journal,a4paper]{IEEEtran}

\usepackage[usenames,dvipsnames]{xcolor}
\usepackage{amsmath,amsthm,graphicx,cite}
\usepackage{epsfig,amsfonts}
\usepackage{graphicx,cite,amssymb,amsmath}
\usepackage{color}
\usepackage{enumitem}
\usepackage[nolist]{acronym}
\usepackage{psfrag}
\usepackage{perso}
\usepackage{xcolor}
\usepackage{blindtext}
\usepackage{relsize}
\usepackage{mathtools}
\mathtoolsset{showonlyrefs}
\usepackage{balance}
\usepackage{placeins}
\newcommand{\nfiles}{n}

\newcommand{\npcached}[1]{w_{#1}}

\newcommand{\npacket}{k}
\newcommand{\codedpacket}{c}

\newcommand{\radius}[1]{r_{#1}}

\newcommand{\dcenter}{d}
\newcommand{\memory}{M}

\newcommand{\library}{\mathcal{F}}
\newcommand{\libraryf}[1]{\mathsf{f}_{#1}}

\newcommand{\hubs}{h}

\newcommand{\probHub}[1]{\gamma_{#1}}
\newcommand{\probFile}[1]{\theta_{#1}}

\newcommand{\isymb}{u}
\newcommand{\isymbv}{\mathbf{\isymb}}

\newcommand{\rosymb}{y}
\newcommand{\rosymbv}{\mathbf{\rosymb}}
\newcommand{\G}{\mathbf{G}}
\newcommand{\Grx}{\tilde{ \mathbf{G}}}
\newcommand{\Expd} [1]{\mathbb{E}[\Delta_{#1}] }

\newcommand{\degreeD}{\Omega}
\newcommand{\Pfail}[1]{P_{F}{#1}}

\newcommand{\Pro}{P}
\newcommand{\rippleset}{\mathscr{R}}
\newcommand{\Ripple}{\mathtt{R}}
\renewcommand{\r}{\mathtt{r}}

\newcommand{\ru}{\r_u}
\newcommand{\Ru}{\Ripple_u}

\newcommand{\cloudset}{\mathscr{C}}
\newcommand{\Cloud}{\mathtt{C}}
\renewcommand{\c}{\mathtt{c}}

\newcommand{\Cu}{\Cloud_u}
\newcommand{\cu}{\c_u}

\renewcommand{\S}[1]{\mathsf{S}_{#1}}

\begin{document}
\begin{acronym}

\acro{LRFC}{linear random fountain code} 
\acro{GEO}{Geostationary Earth Orbit}
\acro{ILP}{Integer Linear Programming}
\acro{LP}{Linear Programming}
\acro{MDS}{maximum distance separable}
\acro{MBS}{Macro Base Station}
\acro{SBS}{Small Base Station}
\acro{Tx}{transmitter}
\acro{mNode}{master node}
\acro{LT}{LT}
\acro{r.v.}{random variable}
\acro{RSD}{robust Soliton distribution} 
\end{acronym}

\title{Caching at the Edge with LT  Codes}
\author{
    \IEEEauthorblockN{Estefan\'ia Recayte, Francisco L\'azaro, Gianluigi Liva\\
    \IEEEauthorblockA{\IEEEauthorrefmark{1}Institute of Communications and Navigation of DLR (German Aerospace Center),
    \\Wessling, Germany. Email:  \{Estefania.Recayte, Francisco.LazaroBlasco,Gianluigi.Liva\}@dlr.de}\\
}
\thanks{This work has been accepted for publication at IEEE ISTC 2018.}
\thanks{\copyright 2018 IEEE. Personal use of this material is permitted. Permission
from IEEE must be obtained for all other uses, in any current or future media, including
reprinting /republishing this material for advertising or promotional purposes, creating new
collective works, for resale or redistribution to servers or lists, or reuse of any copyrighted
component of this work in other works.}
}
\maketitle



\thispagestyle{empty} \pagestyle{empty}

\begin{abstract}
We study the performance of   caching schemes based on LT   under  peeling (iterative) decoding algorithm.  We assume that users ask for downloading content  to multiple cache-aided transmitters.   Transmitters are connected through a backhaul link to a master node while no direct link exists between users and the master node. 
  Each content is fragmented and coded with LT code. Cache placement at each transmitter is optimized  such that  transmissions over the backhaul link is minimized.  We derive a closed form expression for the calculation of the backhaul transmission rate. We compare the performance of a    caching scheme based on LT     with respect to  a caching scheme based on maximum distance separable codes . Finally, we show that caching with \acl{LT} codes behave    as good as caching with maximum distance separable codes. 
\end{abstract}


 \section{Introduction}\label{sec:Intro}

Caching multimedia contents at the network edge is a promising solution in order to mitigate traffic congestion in the backhaul link.  Numerous works  have investigated the potential benefits of caching schemes based on coded content  applied in heterogeneous networks \cite{Liao:2017, ozfatura2018,piemontese2016,bioglio:Globcom2015, Caire:femto}. In literature, 
caching schemes based on \ac{MDS} codes have been  studied  for future wireless networks for maximizing the energy efficiency, minimizing the expected downloading time \cite{piemontese2016} or reducing the amount of data that has to be sent from the core of the network  \cite{bioglio:Globcom2015}. 
 The technique of storing coded fragments of the files significantly outperforms the uncoded approach. Although \ac{MDS} codes represent a benchmark for coding and caching schemes, their practical application is limited by the use of moderate field size due to the high encoding and decoding complexity.
A further disadvantage of \ac{MDS} codes is that their rate is fixed  before the encoding takes place. 

In cache-aided networks a feasible substitute of optimal codes
might be \emph{rateless} codes. Rateless codes are characterized by the fact that the encoder can generate an unlimited amount of coded symbols. In \cite{ASMS:cachingLRFC} it was shown that \ac{LRFC} caching schemes can perform very close to a system based on \ac{MDS} codes. 

In this paper, we analyse the performance of a  caching scheme based on LT codes . In particular, our scheme considers LT codes under  peeling (interative) decoding, which are characterized by a low decoding complexity. We derive a simple   closed form expression for the calculation of the average backhaul transmission rate. The backhaul transmission rate is defined as the number of extra output symbols sent by the \ac{mNode} to a transmitter to serve a user request. We show that caching at the edge with low complexity coding schemes does not significantly penalize  the performance  in terms of backhaul transmission rate.

The remainder of this paper is organized as follows. The system model considered is introduced in Section~\ref{sec:sysmodel}. In Section~\ref{sec:ltcode} the  caching scheme {based on LT codes} is presented. The achievable backhaul rate is derived in Section~\ref{sec:averageRate} while the LT placement optimization problem is formulated in  Section~\ref{sec:placement}. In Section~\ref{sec:results} the numerical results are presented. At last, conclusions are drawn in Section~\ref{sec:Conclusions}. 
 \section{System Model}\label{sec:sysmodel}

\begin{figure}[t]
 \includegraphics[width=\columnwidth]{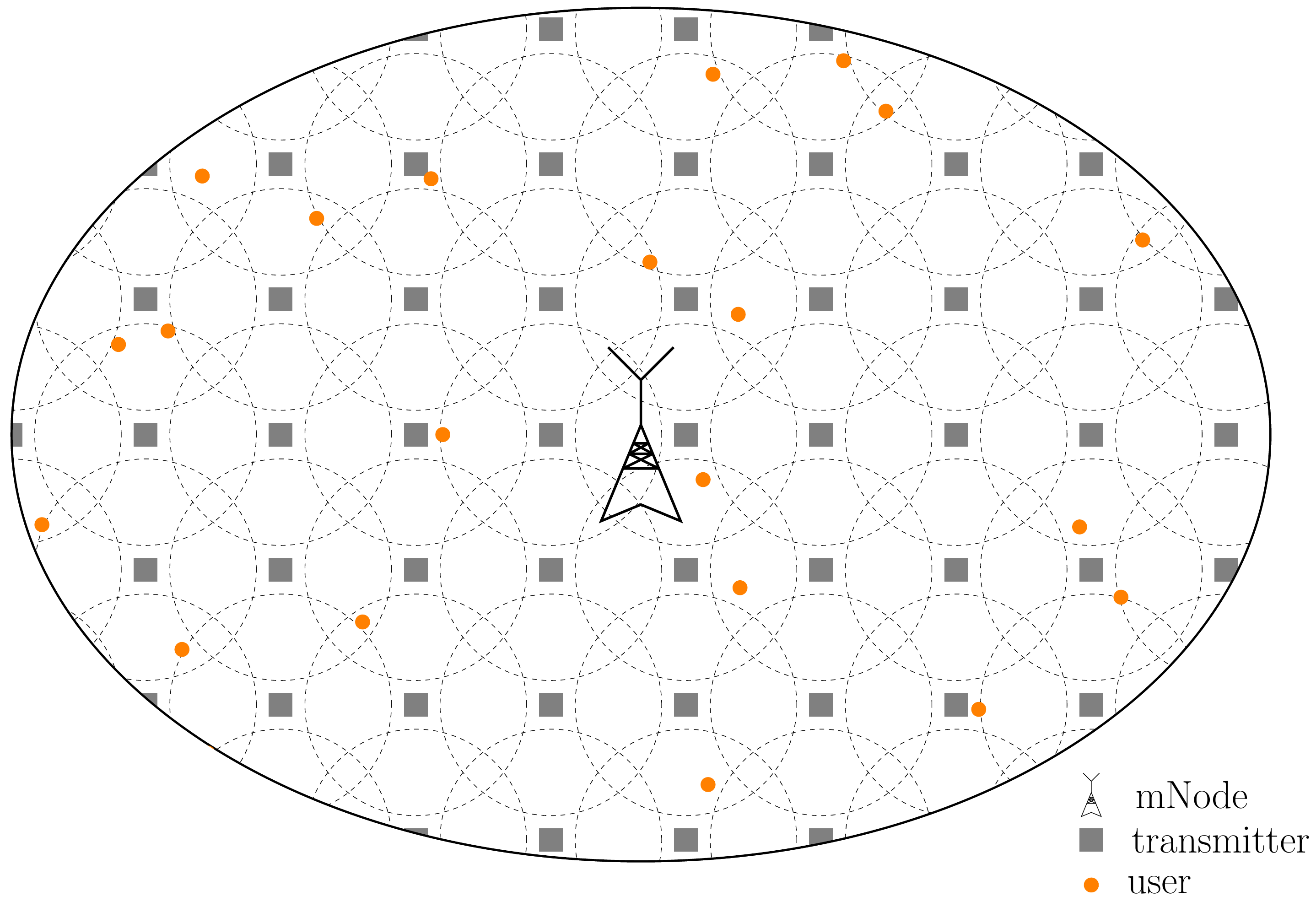}
\centering\caption{System model}
\label{fig:model}
\end{figure}

Let us consider the cache-aided network depicted in Fig.~\ref{fig:model} where users are interested in downloading files from a \ac{mNode}. Transmitters equipped with a local cache
are deployed in the area in order to serve users requests. We assume that each transmitter is connected to the  \ac{mNode} through a wireless backhaul link while no  direct link exists between users and the \ac{mNode}.

 Users send  request to download  files belonging to a library of $\nfiles$ of equal size files $\library = \{\libraryf{1}\, \ldots, \libraryf{\nfiles}\}$. A file $\libraryf{j}$  is requested  with probability $\probFile{j}$, following a Zipf  distribution with shape parameter $\alpha$ such that
\[
\probFile{j} = \frac{1/j^\alpha}{\sum_{i=1}^n 1/i^{\alpha}}\quad   j =1, ...,n.
\]
Each transmitter can store in its local cache $M$ files.  A user can be served by the cached content of  multiple transmitters
 according to his own location. We denote by $\probHub{\hubs}$ the probability that a user is connected to $\hubs$ transmitters.

 We consider a caching scheme {based on LT codes}. The \ac{mNode}, who has access to the whole library $\library$, encodes each file using an LT code. In particular, each file is  fragmented into $\npacket$ input symbols and an LT code is used to generate output (coded) symbols. During the placement phase, each transmitter stores
the same number $\npcached{j}$ of output coded symbols for file  $\libraryf{j}$. Due to the LT construction, each output symbol is generated independently from all other output symbols.
During the delivery phase, users send request for contents belonging to $\library$.  A user initially receives {coded content of the requested file directly from the caches of the transmitters.}  If the user cannot decode the file requested from the cached content then the \ac{mNode} sends supplementary output symbols until the decoding process is declared successful. Extra output symbols are sent via the backhaul link from the \ac{mNode} to one of the transmitters which is serving the users and finally that transmitter forwards the symbols to the user. 
 For simplicity we assume that all transmissions are error-free.

 \section{LT Codes}\label{sec:ltcode}

We consider the use of binary LT codes \cite{luby02:LT} under peeling (iterative) decoding  for the delivery of the different files in the library. Each file is  fragmented into $k$ input symbols,
$ (\isymb_1, \isymb_2, \hdots, \isymb_\npacket )$, which are fed into the LT encoder. The LT encoder generates output symbols
$\mathbf{\codedpacket} = ( \codedpacket_1, \codedpacket_2, \hdots, \codedpacket_{\ell} )$, where the number of outputs symbols $\ell$ can grow indefinitely.
Each output symbol $\codedpacket_i$ is constructed by first sampling a degree $d$ from a degree distribution $\degreeD=(\degreeD_1, \degreeD_2, \degreeD_3, \dots, \degreeD_{d_{\text{max}}}) $, where $d_{\text{max}}$ is the maximum output degree. Next, $d$ distinct symbols are selected uniformly at random among the $k$ source symbols, and their x-or is computed to generate the output symbol.

For fixed $\ell$, LT encoding can be expressed as a vector matrix multiplication
\[
\mathbf{\codedpacket} = \isymbv \G
\]
where $\isymbv$ is the vector of input symbols and $\G$ is the generator matrix of the LT code, being each column of $\G$ associated to an output symbol.

In order to download a file, a user must collect a set of $m \geq \npacket$ output symbols $\rosymbv=(\rosymb_1, \rosymb_2, \hdots, \rosymb_m)$. If we denote by $\mathcal{I} =(i_1, i_2, \hdots, i_m)$ the set of indices corresponding to the $m$ output symbols collected by the receiver we have
\[
\rosymb_r = \codedpacket_{i_r}.
\]
The user attempts decoding by solving the system of equations
\[
\rosymbv = \isymbv \Grx.
\]
where $\Grx$ is the matrix obtained by taking the columns of $\G$ with indices in  $\mathcal{I}$.
In this paper we shall assume that the peeling decoder proposed in \cite{luby02:LT} is used, which is suboptimal but shows a good performance if $k$ is large and the degree distribution is suitably chosen. The peeling decoder is better explained based on a bipartite representation of the LT code. Decoding consists of initially marking all input symbols as unresolved and then carrying out $k$ decoding stages. At every decoding stage the decoder looks for a degree one output symbol. If a degree one output symbol $\rosymb$ is found, its only neighbor $\isymb$ is resolved, and all edges attached to $\isymb$ are erased from the bipartite graph, which reduces the degree of the neighbors of $\isymb$. If no degree one output symbol is found a decoding failure is declared. After $k$ decoding stages without a failure all $k$ input symbols are recovered and decoding succeeds.

The performance of the peeling decoder was analyzed in \cite{Karp2004,shokrollahi2009theoryraptor} using dynamic programming. In the following we shall make use of the notation used in \cite{lazaro:inact} to describe the peeling decoder.
Let us define  the output ripple (or simply ripple) as the set of output symbols of degree 1, denoted $\rippleset$, and let us define the cloud as the set of output symbols of degree $d\geq 2$, denoted $\cloudset$. Furthermore, we shall denote the random variables associated to the cardinality of the {cloud and ripple} by   and  $\Cloud$ and $\Ripple$, respectively, and by $\c$ and $\r$ their realizations. Moreover, we shall use the subscript $u$ to represent the number of unresolved input symbols, so that $\Cu$, for example, represents the cardinality of the cloud when $u$ input symbols are unresolved.
In \cite{Karp2004,shokrollahi2009theoryraptor} the LT decoder is modelled as a finite state machine. In particular, given the absolute receiver overhead $\delta=m-k$, the decoder is modelled as a finite state machine with state
\[
\S{u}:=(\Cu, \Ru ).
\]
Based on this model, a recursion is derived that allows to obtain  $\Pr \{ \S{u-1}=(\c_{u-1}, \r_{u-1}) \}$ as a function of ${\Pr \{ \S{u}=(\cu, \ru )\}}$. This recursion allows deriving the exact probability of the decoder being at state $\S{u}:=(\cu, \ru )$ for $u=k,k-1,\hdots, 1$. By observing that decoding fails whenever the ripple is empty ($\ru=0$), the   probability of decoding failure can be obtained as
\begin{align}\label{pfailure}
\Pfail(\delta) = \sum_{u=1}^{\npacket} \sum_{\cu} \Pr\{\S{u} =(\cu,0) \}
\end{align}
where we observe that $\Pfail(\delta)=1$ for $\delta < 0$.
\section{Average Backhaul Rate} \label{sec:averageRate}

\subsection{Overhead Average} \label{susec:DecProb}
Let  $\Delta$ be the random variable associated to the number of LT output symbols  in excess to $k$ that { a user}  needs to successfully decode the requested content. We denote with $\delta$  a realization of  $\Delta$.
We can calculate the expected value of $\Delta$, i.e. the average overhead, as follows
 \begin{align}\label{eq:delta_av}
\Expd{} & = 
	      \sum_{\delta=1}^{\infty} \delta   \cdot \Big[\Pfail(\delta-1)-\Pfail(\delta) \Big]\\
	    & = \sum_{\delta= 0}^{\infty} (\delta+1)\cdot \Pfail(\delta)  - \sum_{\delta= 0}^{\infty} \delta  \cdot \Pfail(\delta)	\\
	     &= \sum_{\delta=0}^{\infty}  \Pfail(\delta).
\end{align}

\subsection{Backhaul Transmission  Rate}

The average backhaul transmission  rate is calculated as  proposed in \cite{ASMS:cachingLRFC}. 

The total number of LT cached symbols of the file requested that a user receives  from the transmitters is modelled by the \ac{r.v.} $Z$, denote by $z$ the realization of $Z$. Let also denote by $H$ the \ac{r.v.} associated to the number of transmitters connected to the user of interest, and by $h$ its realization. The index of the file requested by a user is modelled by the \ac{r.v.} $J$ and with $j$ we indicate its realization. We have
\begin{align}\label{eq:Z}
P_{Z|J,H}(z | j, h) = \begin{cases}
                    1  & \mbox{if } z= \npcached{j} \, h\\
                    0  & \mbox{otherwise}
                    \end{cases}
\end{align}
where we recall that  $\npcached{j}$ stands for the number of coded symbols from file $j$ stored in every transmitter. The probability mass function of $Z$ is
\begin{align}\label{eq:z_2}
    P_Z(z)    &= \sum_j \sum_h  P_{Z|J,H}(z | j, h) \cdot P_J(j) \cdot P_H(h)  \\
   & =  \sum_j \sum_h \probFile{j} \, \probHub{\hubs} \,  P_{Z|J,H}(z | j, h).
\end{align}

We are interested in deriving the distribution of the number of output symbols which have to be sent over the backhaul link to serve the request of a user, which we denote by random variable $T$. If we condition $T$ to $Z$, it is easy to see how the probability of $T=t$ corresponds to the probability that decoding succeeds when the user has received exactly $t$ output symbols from the backhaul  link in excess to the  $z$ output symbols it received from the transmitters through local links, that is, when $m=z+t$.
\newline
 In order to derive $P_{T|Z}(t|z)$ we shall distinguish two cases.

 If $z \leq k$, then
\begin{align}\label{eq:t_cases1}
&  P_{T|Z}(t|z) =
\begin{cases}
 \displaystyle    \Pfail(z-k+t-1) - \Pfail(z-k+t ) \mkern10mu   \mbox{if } t \geq 0   \\
  0  \mkern255mu  \mbox{otherwise.}
\end{cases}
\end{align}

If $z>k$, then
\begin{align}\label{eq:t_cases2}
&  P_{T|Z}(t|z) =
\begin{cases}
 \displaystyle  1 - \Pfail(z-k)   \mkern165mu   \mbox{if } t = 0,& \\  \displaystyle  \Pfail(z-k+t-1) - \Pfail(z-k+t )  \mkern10mu \mbox{if t} > 0,& \\
                   0    \mkern255mu  \mbox{otherwise.} &
  \end{cases}
\end{align}

As in \cite{ASMS:cachingLRFC},  the expectation $ \mathbb{E}[T]$ is given by
\begin{align}\label{eq:t}
  \mathbb{E}[T] = &\sum_t t \Bigg( \sum_{z=0}^{k} \Pro_{T|Z}(t|z) \Pro_Z(z) \\
 &  + \sum_{z=k+1}^{\infty} \Pro_{T|Z}(t|z) \Pro_Z(z)  \Bigg).  \\
\end{align}

Let us define $\bar{T}_1$ and $\bar{T}_2$ as
\begin{align}
  \bar{T}_1 :&=  \sum_t t  \sum_{z=0}^{k}\Pro_{T|Z}(t|z) \Pro_Z(z) \\
  \bar{T}_2 :&=  \sum_t t \sum_{z=k+1}^{\infty} \Pro_{T|Z}(t|z) \Pro_Z(z)
\end{align}
so that
\begin{align} \label{eq:T1T2}
 \mathbb{E}[T] = \bar{T}_1 + \bar{T}_2.
\end{align}

If we introduce the variable change $\delta = z-k+t$ in the expression of $\bar{T}_1$, we obtain
\begin{align}
  \bar{T}_1 & =
  \sum_{z=0}^{k}  \Pro_Z(z)  \sum_{\delta =z-k}^{\infty} (\delta  -z+k )\Big[\Pfail(\delta - 1)-\Pfail(\delta )\Big]   \\
  & \stackrel{(\mathrm{a})}{=}  \sum_{z=0}^{k} \Pro_Z(z)  \sum_{\delta  = 0}^{\infty} (\delta  -z+k )\Big[\Pfail(\delta- 1)-\Pfail(\delta )\Big] \\
  &=    \sum_{z=0}^{k}  \Pro_Z(z) \Bigg( \sum_{\delta=0}^{\infty} \delta \Big[\Pfail(\delta - 1)-\Pfail(\delta )\Big]  \\
 &  +  (k-z) \sum_{\delta=0}^{\infty}  \Big[\Pfail(\delta - 1)-\Pfail(\delta )\Big]  \Bigg)  \\
  &\stackrel{(\mathrm{b})}{=}    \sum_{z=0}^{k} \Pro_Z(z)   \Big(  \Expd{} +  k-z \Big)   \\
  &= ( \Expd{} +  k)  \Pr\{Z \leq k\}   -  \sum_{z=0}^{k} z   \Pro_Z(z)  \label{eq:T1}
\end{align}
where equality $(\mathrm{a})$ is due to $[\Pfail(\delta - 1)-\Pfail(\delta)]=0$ for $\delta < 0$ and equality $(\mathrm{b})$ is due to
\[
\sum_{\delta=0}^{\infty} [\Pfail(\delta - 1)-\Pfail(\delta )]  = 1.
\]

Introducing the same variable change in the expression of $\bar{T}_2$ we have
\begin{align}
  \bar{T}_2 &=   \sum_{z=k+1}^{\infty}  \Pro_Z(z)  \sum_{\delta=z-k}^{\infty} (\delta -z+k )\Big[\Pfail(\delta - 1)-\Pfail(\delta )\Big] \\
  & = \sum_{z=k+1}^{\infty}  \Pro_Z(z)  \Bigg( \sum_{\delta=z-k}^{\infty} \delta \Big[\Pfail(\delta - 1)-\Pfail(\delta )\Big] \\
  & + \sum_{\delta=z-k}^{\infty} (k-z) \Big[\Pfail(\delta - 1)-\Pfail(\delta )\Big] \Bigg).\label{eq:upT}
\end{align}
Let us rewrite   \eqref{eq:upT} as follow
\begin{align}
  \bar{T}_2 &=   \sum_{z=k+1}^{\infty} \Pro_Z(z) \Big(\bar{T}_{21}(z) +  \bar{T}_{22}(z)\Big)   \label{eq:T2}
 \end{align}
 where
\begin{align}
&  \bar{T}_{21}(z) =
  \sum_{\delta=z-k}^{\infty} \delta  \Big[\Pfail(\delta - 1)-\Pfail(\delta )\Big] \\
=&  \sum_{\delta=0}^{\infty} \delta \Big[\Pfail(\delta - 1)-\Pfail(\delta )\Big] - \sum_{\delta=0}^{z-k-1} \delta \Big[\Pfail(\delta - 1)-\Pfail(\delta )\Big] \\
=&  \sum_{\delta= 0}^{\infty} \Pfail(\delta)  - \Bigg[ \sum_{\delta=0}^{z-k-1}  \Pfail(\delta) -(z-k)\Pfail(z-k-1)  \Bigg]\\
=&  \Expd{} - \sum_{\delta=0}^{z-k-1}  \Pfail(\delta) +(z-k) \Pfail(z-k-1) \label{eq:T21}
\end{align}
and
\begin{align}
\bar{T}_{22}(z)
=& \sum_{\delta=z-k}^{\infty} (k-z) \Big[\Pfail(\delta - 1)-\Pfail(\delta )\Big]  \\
=& (k-z) \Bigg\{ \sum_{\delta=0}^{\infty}  \Big[\Pfail(\delta - 1)-\Pfail(\delta )\Big] \\
-& \sum_{\delta=0}^{z-k-1}  \Big[\Pfail(\delta - 1)-\Pfail(\delta )\Big] \Bigg\}\\
=&  (k-z) \Big\{1 - \Big[1 -\Pfail(z-k-1)\Big] \Big\}  \\
=&  (k-z)  \Pfail(z-k-1)     \label{eq:T22}.
\end{align}
By inserting \eqref{eq:T21} and  \eqref{eq:T22} in \eqref{eq:T2} and sum we obtain
\begin{equation}
\bar{T}_2 =   \sum_{z=k+1}^{\infty}  \Pro_Z(z) \Big[ \Expd{}- \sum_{\delta=0}^{z-k-1}  \Pfail(\delta)   \Big ]\label{eq:T2f}.
\end{equation}
Finally, if we replace \eqref{eq:T1} and \eqref{eq:T2f} in \eqref{eq:T1T2}, the expression of the average backhaul transmission rate becomes
\begin{align} \label{eq:ET}
&\mathbb{E}[T]   =
   \Big( \Expd{} +  k \Big)  \Pr\{Z \leq k\}   -  \sum_{z=0}^{k} z   \Pro_Z(z)   \\
 &  +  \sum_{z=k+1}^{\infty}\Pro_Z(z) \Big[ \Expd{}- \sum_{\delta=0}^{z-k-1}  \Pfail(\delta)  \Big] \\
 = & \Expd{} +    \sum_{z=0}^{k}(k- z)\Pro_Z(z)
  - \sum_{z=k+1}^{\infty}  \Pro_Z(z) \Bigg[  \sum_{\delta=0}^{z-k-1}  \Pfail(\delta)\Bigg].\\
\end{align}
The last term in \eqref{eq:ET} is always non negative. Hence, we can 
upper bound \eqref{eq:ET} as 
\begin{equation}
 T_{\text{UP}}  =
  \Expd{} +    \sum_{z=0}^{k}(k- z)\Pro_Z(z) \geq  \mathbb{E}[T]
   .\\
\label{eq:ETbound}
\end{equation}

\section{  LT  Placement Optimization}\label{sec:placement}

The goal of the  LT placement optimization  problem is to define the number of output coded symbols per file that has to be stored  in the transmitters' caches such that during the delivery phase the average backhaul link is minimized.

 The LT placement optimization  problem can be formulated as follows
\begin{align} \label{eq:optProblem1}
\underset{\npcached{1}, \dots, \npcached{\nfiles}}  \min &  \quad    T_{\text{UP}} \\ \label{eq:optProblem2}
 \text{s.t. }& \quad \sum_{j=1}^{\nfiles}   \npcached{j}   = \memory \npacket  \\
&  \label{eq:optProblem4} \quad     \npcached{j} \in  \mathbb{N} \ \quad j = 1, \ldots, \nfiles.
\end{align}
where  \eqref{eq:optProblem2}  limits total number of stored LT coded symbols to the size cache capacity, 
while  \eqref{eq:optProblem4}   accounts for the discrete nature of the optimization variable.

The evaluation on the exact expression  $\Expd{}$ is more complex than the evaluation of the upper bound. We hence opt for the minimization of the upper bound, comforted by the numerical results presented next.

A suboptimal solution of the problem in \eqref{eq:optProblem1}-\eqref{eq:optProblem4} can be obtained by using linear programming methods  when the constraint   \eqref{eq:optProblem4} is relaxed.

\section{Results}\label{sec:results}

\begin{figure}[t]
 \includegraphics[width=\columnwidth]{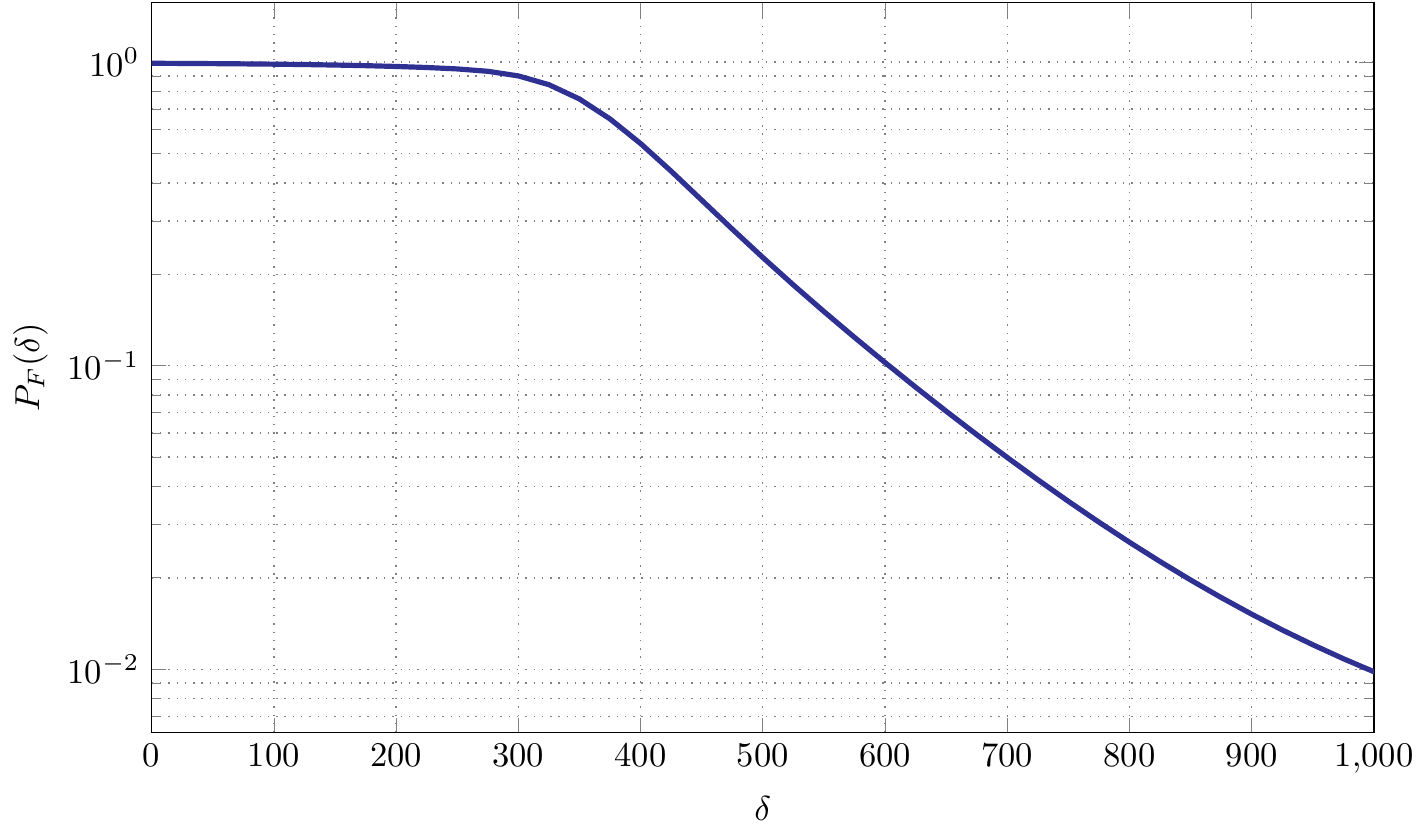}
\centering \caption{Failure probability in function of overhead $\delta$ for a robust soliton distribution  with parameters $c = 0.05$ and $d = 3$ for k = 10000 under  peeling iterative decoding.}
\label{fig:P_fail}
\end{figure}

 \begin{figure}[t]
 \includegraphics[width=\columnwidth]{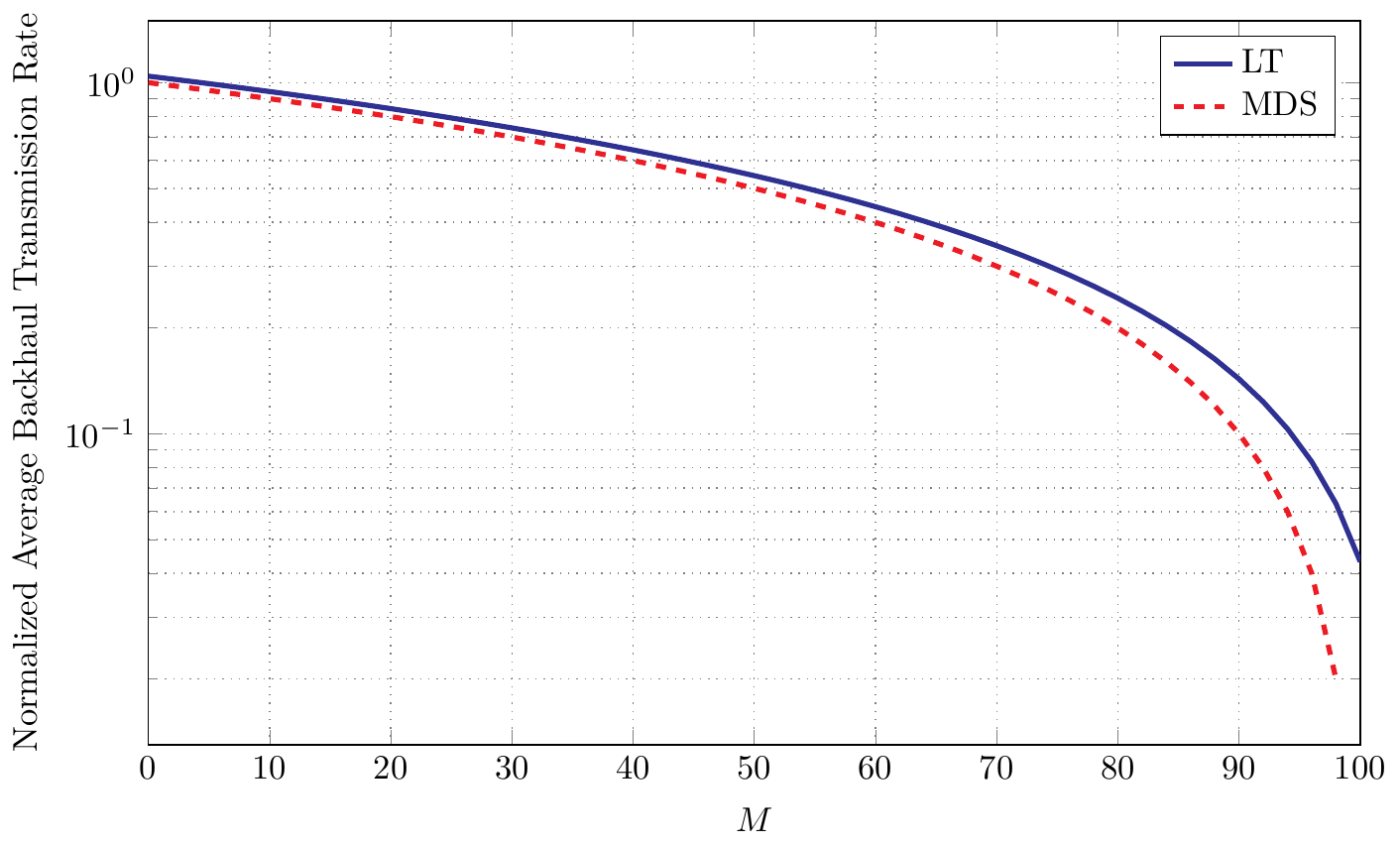}
 \centering \caption{Normalized average backhaul transmission rate as a function of memory size $M$ for \ac{LT} and \ac{MDS} codes  given $n=100$, $k=10000$, $\alpha=0 $ and $\gamma_1=1$.}
 \label{fig:R_uniform}
\end{figure}

We are interested in evaluating the normalized backhaul transmission rate defined as $\mathbb{E}[T] /k$, {where} $k$ the number of fragments of a file. Given the connectivity $\gamma_h$ and the requested $\theta_j$  probability distribution for each scenario, we optimize   the number of fragments $w_j$ to be cached during the placement phase  by solving the optimization problem $\eqref{eq:optProblem1}$-$\eqref{eq:optProblem4}$.  Finally, given $w_j$   we calculate $\mathbb{E}[T] /k$ from \eqref{eq:ET}.

We consider the system described in Section \ref{sec:sysmodel} which is illustrated in Fig.~\ref{fig:model}. We assume users to be uniformly distributed and transmitters have an area of coverage of $\radius{}=60$ m. We further assume that  each transmitter is deployed according to a uniform two dimensional grid, with spacing  $\dcenter = 45$ m. A user can be served by multiples caches due to the fact that the coverage area of transmitters partially overlap. In particular, a user can be served by $h$ caches with probability $\gamma_h$.
With simple geometrical calculations  we can derive
the following connectivity distribution
\begin{equation}\label{eq:gammadis}
 \gamma_1 = 0.2907,  \,  \gamma_2 =0.6591, \, \gamma_3 = 0.0430, \, \gamma_4 =0.0072.
\end{equation}

 In all setups, we consider that each file is fragmented in ${k=10000}$ input symbols.
We assume that the \ac{mNode} implements an LT code   characterized by a \ac{RSD} \cite{luby02:LT} with parameter $c = 0.05  $ and $d = 3$, where $c$ and $d$ have been chosen so that  the average overhead is minimized. The probability of decoding failure $P_F(\delta)$ was derived as explained in Sec. \ref{sec:ltcode}.

In Fig.~\ref{fig:P_fail}  the  probability of decoding failure as a function of $\delta$ for  \ac{RSD} under iterative peeling decoding is plotted. From \eqref{eq:delta_av}, we obtained that the overhead average  is $\Expd{}= 431,95$.

In our first setup, we study the normalized average backhaul transmission rate as a function of the cache size $M$ when the library  cardinality is $n=100$. We consider that each file has the same probability to be requested, i.e. Zipf distribution with $\alpha=0$.
We assume that each user can be connected to only one transmitter, i.e. $\gamma_1=1$.
In Fig.~\ref{fig:R_uniform}, we can  observe that the \ac{LT} coded caching scheme performs close to the \ac{MDS} coded caching scheme. The cost in terms of average transmission rate for using a \ac{LT} code is only the 4.32\% which coincides with the average overhead of the LT code.

 \begin{figure}[t]
 \includegraphics[width=\columnwidth]{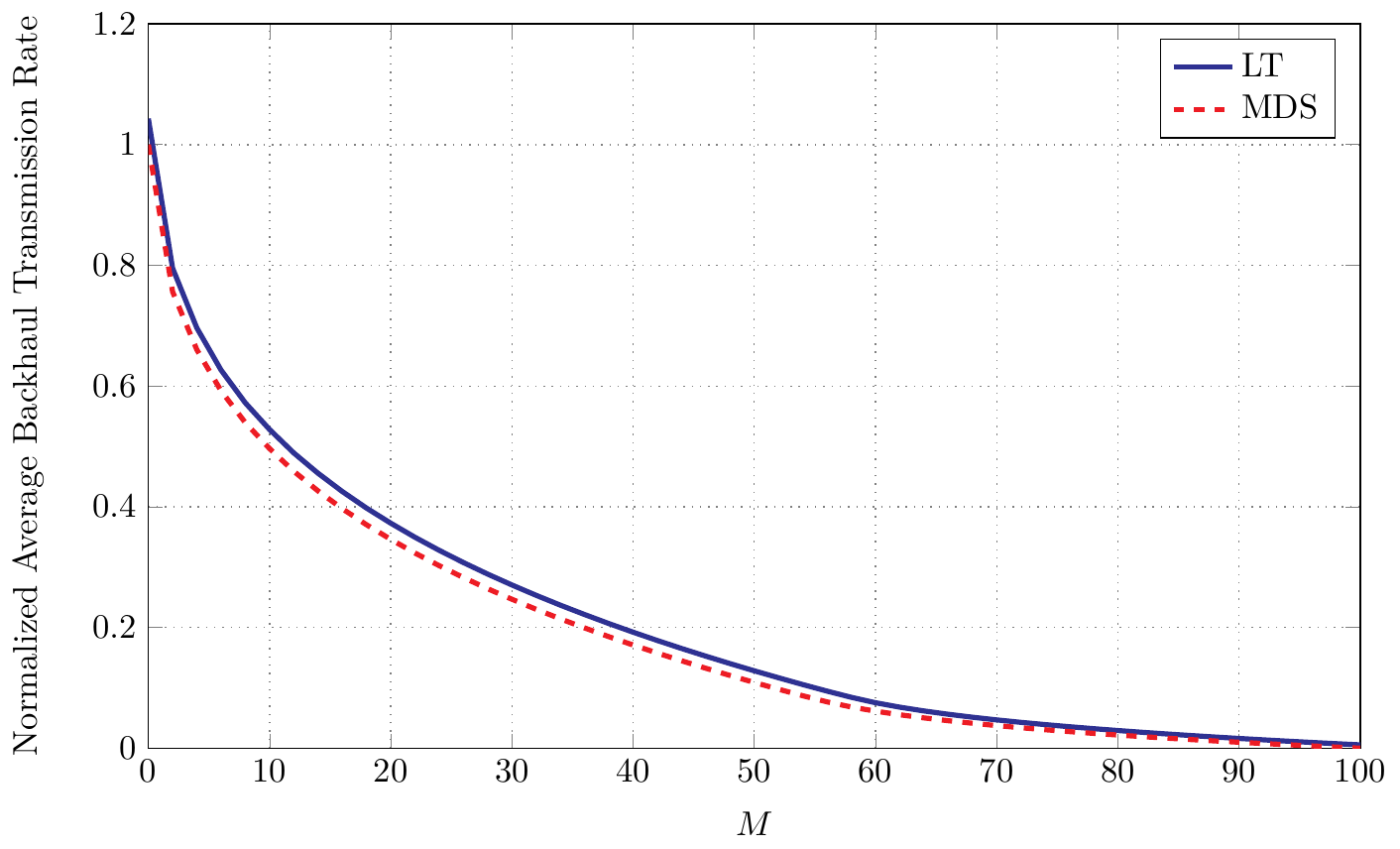}
 \centering \caption{Normalized average backhaul transmission rate as a function of memory size $M$ for \ac{LT} and \ac{MDS} codes  given $n=100$, $k=10000$, $\alpha=0.8 $ and $\gamma_1=0.2907, \gamma_2=0.6591, \gamma_3=0.0430 \text{ and } \gamma_4=0.0072$.}
 \label{fig:R_M}
\end{figure}

In our second setup, we consider the connectivity distribution given in   \eqref{eq:gammadis}. We also assume that files are requested according to a  Zipf distribution with ${\alpha = 0.8}$.
In Fig. \ref{fig:R_M} the normalized average backhaul transmission rate is shown as a function of the cache size $M$. We can see that the \ac{LT} caching scheme is comparable to the benchmark \ac{MDS} scheme. Furthermore, as the cache size increases, the difference between the two approaches becomes negligible. 

\begin{figure}[t]
 \includegraphics[width=\columnwidth]{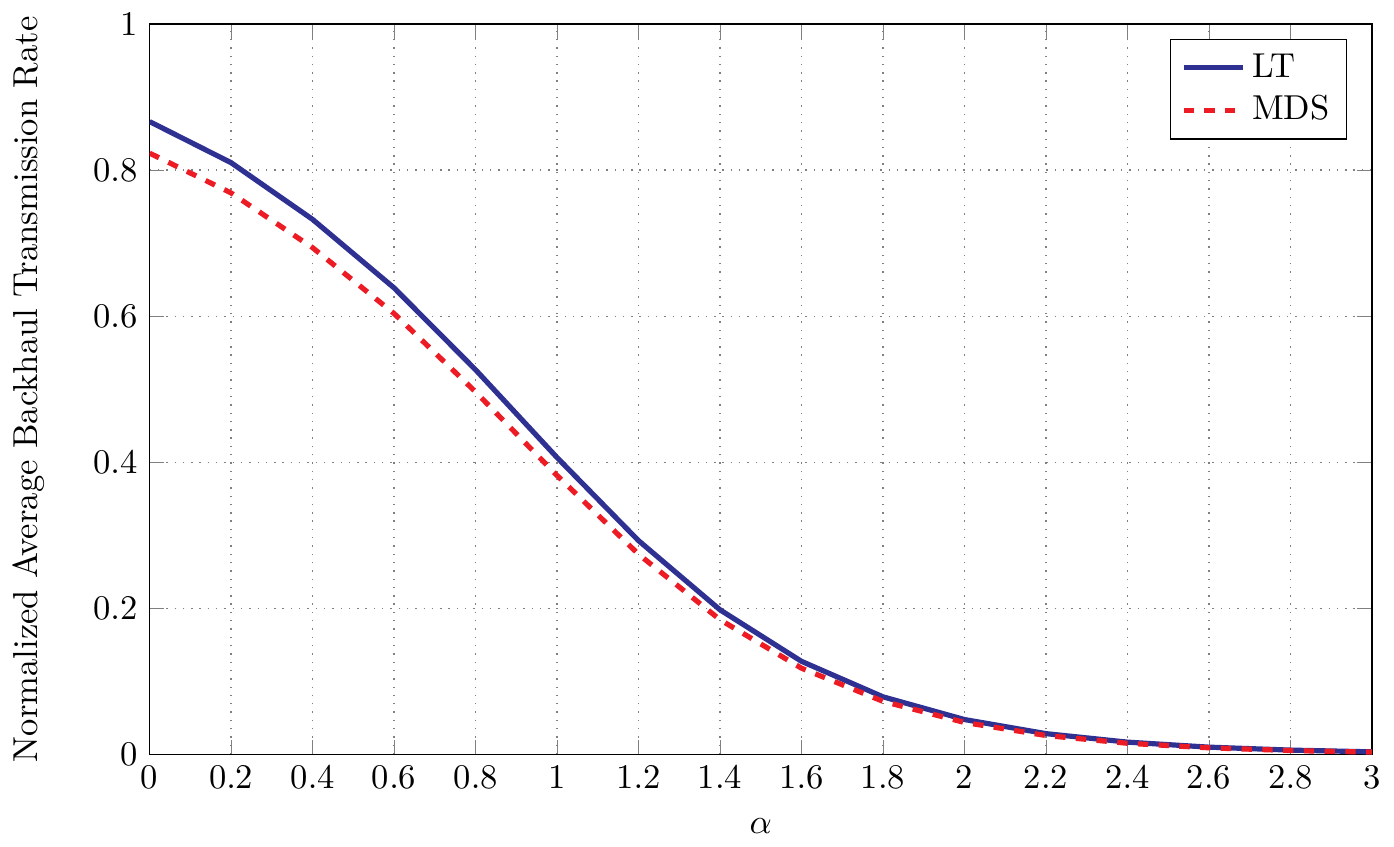}
\centering \caption{Normalized average backhaul transmission rate as a function of the file parameter distribution $\alpha$ for \ac{LT} and \ac{MDS}  given $n=100$, $k=10000$, ${M=10}$ and $\gamma_1=0.2907$, $\gamma_2=0.6591$, $\gamma_3=0.0430$, $\gamma_4=0.0072$.}
\label{fig:R_alpha}
\end{figure}

In our last case, we consider the same  connectivity $\gamma_h$   as the previous setup. We assume to have a fixed memory size ${M=1}0$ and library size $n=100$ files. In Fig. \ref{fig:R_alpha} the normalized average backhaul transmission rate is shown as a function of the shape parameter $\alpha$.
As expected,  for  $\alpha=0$, i.e. when files are equiprobable, the \ac{MDS} scheme outperforms the scheme based on LT codes. However as $\alpha$ increases, the performance of the LT scheme approaches that of the \ac{MDS} scheme.

\section{Conclusions}\label{sec:Conclusions}

We considered a scheme were coded content is cached at the edge in a heterogeneous network, with the aim of minimizing the number of transmission from the core of the network during the delivery phase. In particular, we considered the use of LT codes under peeling (iterative) decoding. We delivered an analytical expression of the average backhaul transmission rate. We also formulated the    optimization problem related to the placement phase of the LT code based caching scheme. Finally, we compared the performance of the LT scheme with that of an optimal but impractical \ac{MDS} scheme. Our simulation results indicate that the performance of LT codes approaches that of the optimal scheme, but exhibiting a much lower decoding complexity.

\balance

\bibliographystyle{IEEEtran}



\end{document}